\documentclass[aps,twocolumn,showpacs,superscriptaddress]{revtex4-1}
\usepackage{amsmath,amssymb}
\usepackage{float}
\usepackage[utf8]{inputenc}
\usepackage[plainpages=false]{hyperref}
\usepackage{amssymb}
\usepackage{mathrsfs}
\usepackage{graphicx}
\usepackage[caption=false]{subfig}
\usepackage{mathtools}
\usepackage[ruled, linesnumbered]{algorithm2e}
\usepackage{appendix}

\begin{document}

\title{Generating photons from vacuum with counter rotating wave interaction}

\author{Hai-Jing Song}

\affiliation{National Innovation Institute of Defense Technology, AMS, Beijing 100071, China.}
	
\affiliation{Institute of Physics, Beijing National Laboratory for
		Condensed Matter Physics, Chinese Academy of Sciences, Beijing
		100190, China}
	
\affiliation{School of Physical Sciences, University of Chinese	Academy of Sciences, Beijing 100049, China}

\author{Qi-Kai He}

\affiliation{Institute of Physics, Beijing National Laboratory for
	Condensed Matter Physics, Chinese Academy of Sciences, Beijing
	100190, China}

\affiliation{School of Physical Sciences, University of Chinese
	Academy of Sciences, Beijing 100049, China}

\affiliation{Hangzhou AlgoCity Technology Co., Ltd., Hangzhou 311121, China}

\author{Zheng An}

\affiliation{Institute of Physics, Beijing National Laboratory for
		Condensed Matter Physics, Chinese Academy of Sciences, Beijing
		100190, China}
	
\affiliation{School of Physical Sciences, University of Chinese	Academy of Sciences, Beijing 100049, China}
	
\affiliation{Quantum Science Center of Guangdong-Hong Kong-Macao Greater Bay Area (Guangdong), Shenzhen 518045, China}

\author{D.~L. Zhou}

\email{zhoudl72@iphy.ac.cn}

\affiliation{Institute of Physics, Beijing National Laboratory for
  Condensed Matter Physics, Chinese Academy of Sciences, Beijing
  100190, China}

\affiliation{School of Physical Sciences, University of Chinese
  Academy of Sciences, Beijing 100049, China}

\date{\today}

\begin{abstract}
  		
  We propose a bang-bang control scheme to enhance photon generation from the vacuum via the counter-rotating wave (CRW) interaction, and develop a pruning greedy algorithm (PGA) to identify the optimal control sequence. Our numerical results demonstrate that the maximum number of photons generated within a given evolution time is increased by several orders of magnitude compared with that achieved by continuous activation of the CRW interaction.
  
\end{abstract}


\maketitle

\section{Introduction~\label{sec:intro}}
Recently, atom–photon interactions have been explored in the ultrastrong coupling regime, where the coupling is so strong that the rotating wave approximation (RWA)~\cite{1443594, Burgarth2024taming} breaks down. In this regime, the counter-rotating wave (CRW) interaction becomes significant and gives rise to new phenomena, including inelastic single-photon scattering with an atom in a one-dimensional supercavity or waveguide~\cite{ Zhou2008, He_singlephoton, Wang2022Phase, liu2022Tunable, zhou2023Chiral},
 a photon-populated ground state in the Rabi model~\cite{PhysRev.49.324, Minganti2024Phonon},  and multiphoton quantum Rabi oscillations~\cite{ Mukhopadhyay2024quantum, Doicin2025Multilevel}. The CRW effect has been observed in various systems, such as circuit QEDs~\cite{NDH+2010, liu2023quantum, Mojtaba2025Engineering}, spiropyran molecules~\cite{SHGE2011}, and trapped ions~\cite{PhysRevX.8.021027}, highlighting the growing interest in investigating CRW interactions.

The Rabi model~\cite{PhysRev.49.324, PhysRev.51.652},which describes the coupling between a single-mode cavity and a two-level atom, provides a standard framework for studying the CRW interaction. Its Hamiltonian (with $\hbar = 1$ ) is given by
\begin{equation}
  \hat{H} = \omega_{c} \hat{a}^{\dagger} \hat{a} + \dfrac{\omega_{a}}{2} 
  \hat{\sigma}_{z} + g \hat{\sigma}_{x} (\hat{a}^{\dagger} + \hat{a}),
  \label{eq:rabi}
\end{equation}
where $\hat{a}^{\dagger}$ ($\hat{a}$) is the photon creation
(annihilation) operator for the cavity with frequency
$\omega_{c}$, $\hat{\sigma}_{-} = | g \rangle \langle e |$
($\hat{\sigma}_{+} = \hat{\sigma}^{\dagger}_{-}$) are the atomic lowering and raising operators, $|g\rangle$ and $|e\rangle$ denote the ground and excited states of the two-level atom, 
$\hat{\sigma}_{z}=\hat{\sigma}_{+}\hat{\sigma}_{-}-\hat{\sigma}_{-}\hat{\sigma}_{+}$, 
$\hat{\sigma}_{x}=\hat{\sigma}_{+}+\hat{\sigma}_{-}$, $\omega_{a}$ is the atomic transition frequency, and $g$ is the atom–cavity coupling strength. The interaction term can be decomposed as
\begin{equation}
  \hat{H}_{\rm{int}} = \hat{H}^{\rm{RW}}_{\rm{int}} + \hat{H}^{\rm{CRW}}_{\rm{int}},
  \label{eq:hint}
\end{equation}
with $\hat{H}^{\rm{RW}}_{\rm{int}} = g (\hat{\sigma}_{+} \hat{a} +
\hat{a}^{\dagger} \hat{\sigma}_{-})$ the rotating-wave term and
\begin{equation}
  \hat{H}^{\rm{CRW}}_{\rm{int}} = g (\hat{\sigma}_{+} \hat{a}^{\dagger} + \hat{a} \hat{\sigma}_{-})
  \label{eq:int_CRW}
\end{equation}
the CRW term.

Within the RWA, the CRW term is omitted, yields the Jaynes–Cummings model ~\cite{1443594} whose ground state is $| 0 \rangle \otimes | g \rangle$ with $| 0 \rangle$ denoting the cavity vacuum. When the CRW term is included, the ground state of the Rabi model acquires nonzero photons, suggesting that the CRW interaction can act as a photon generator. However, even in the ultrastrong coupling regime, simply letting the Rabi model evolve freely from the vacuum generates only a few photons. Moreover, achieving extremely large coupling strengths remains an experimental challenge~\cite{Blais2021Circuit, Tay2025Multimode, de2025There}. Therefore, generating more photons from the initial vacuum $|0\rangle\otimes|g\rangle$ using the CRW interaction remains an open challenge.

The Time-tunable qubit-resonator coupling has been realized in experiments~\cite{Peng2023Deterministic, Akbari2025Floquet}.  In particular,  a “symmetry-protected charge qubit” has been designed, allowing the coupling to a superconducting resonator to be tuned from the extremely weak to the ultrastrong regime while keeping the qubit frequency unchanged~\cite{he2019Tunable}. Such time-dependent control of the coupling strength is a key ingredient for the protocol proposed here to generate photons via the CRW interaction.

In this work, our aim is to enhance the generation of photons from the vacuum by employing bang-bang (on/off switching) control of the CRW interaction~\cite{artstein1980discrete, 
sonneborn1964the, PhysRevLett.82.2417, Innocenti2020Ultrafast, Sweilam2021Optimal}. As we will show, the maximum number of photons generated within a fixed time interval can be increased by several orders of magnitude compared with the case where the CRW interaction is continuously active. To this end, we develop a pruning greedy algorithm (PGA) to search for optimal control sequences. We also analyze several new phenomena, including the appearance of photon number plateaus, the dependence of the maximum photon number on the atomic frequency, and the upper limit of the photon number for a fixed total on-time of the CRW interaction.

The paper is organized as follows. In Sec.~\ref{sec:model},
we introduce our model and method. With the help of the PGA proposed in 
Appendix~\ref{app:PGA}, we obtain the optimal control sequence of the CRW 
interaction and observe several new phenomena in Sec.~\ref{sec:bangbang}. In 
Sec.~\ref{sec:conclusion}, we draw the conclusions.

\section{Model and method \label{sec:model}}

In the bang-bang protocol, when the CRW interaction is switched on, the time evolution is governed by the Rabi Hamiltonian~\eqref{eq:rabi}; when it is switched off, the evolution is governed by the free Hamiltonian
\begin{equation}
\hat{H}_{0} = \omega_{c} \hat{a}^{\dagger} \hat{a} + \dfrac{\omega_{a}}{2} \hat{\sigma}_{z}.
\label{eq:h0}
\end{equation}

The system is initially prepared in the vacuum state
$| 0 \rangle \otimes | g \rangle$. Both the Rabi Hamiltonian $H$ and the free Hamiltonian $\hat{H}_{0}$ conserve parity, where the parity is defined as the even–odd parity of the total excitation number
$\hat{N}_{e}=\hat{a}^{\dagger}\hat{a}+|e\rangle\langle e|$. Since the initial state has even parity, the Hilbert space can be truncated to the even-parity subspace~\cite{Liu2024Parity}. As we are interested only in the photon number rather than the atomic state, it is convenient to rewrite Eq.~\eqref{eq:rabi} as an effective Hamiltonian~\cite{qin2024Quantum}
\begin{equation}
  \hat{H}_{\rm{eff}} = \omega_{c} \hat{b}^{\dagger} \hat{b} - 
  \dfrac{\omega_{a}}{2} (-1)^{\hat{b}^{\dagger} \hat{b}} + g 
  (\hat{b}^{\dagger} + \hat{b}).
  \label{eq:heff}
\end{equation}
Similarly, the effective version of Eq.~\eqref{eq:h0} is
\begin{equation}
  \hat{H}^{\rm{eff}}_{0} = \omega_{c} \hat{b}^{\dagger} \hat{b} - 
  \dfrac{\omega_{a}}{2} (-1)^{\hat{b}^{\dagger} \hat{b}},
  \label{eq:h0eff}
\end{equation}
whose eigenstates are $| m \rangle$ satisfying $\hat{b}^{\dagger} \hat{b} | m \rangle = m | m \rangle$. It is worth noting that the average photon numbers obtained from the evolution under $\hat{H}$ and $\hat{H}_{\rm{eff}}$ are identical, i.e.,
$\langle \hat{a}^{\dagger} \hat{a} \rangle = \langle \hat{b}^{\dagger} \hat{b} \rangle$.

Evolution under the free Hamiltonian $\hat{H}^{\rm{eff}}_{0}$ only changes the relative phase between components of the quantum state and leaves the average photon number unchanged. Therefore, the variation of the photon number is determined solely by the dynamics under $\hat{H}_{\text{eff}}$, motivating us to examine how $\hat{H}_{\rm{eff}}$ drives changes in the average photon number.

\begin{figure}[!htbp]
  \centering
  \includegraphics[width=0.45\textwidth]{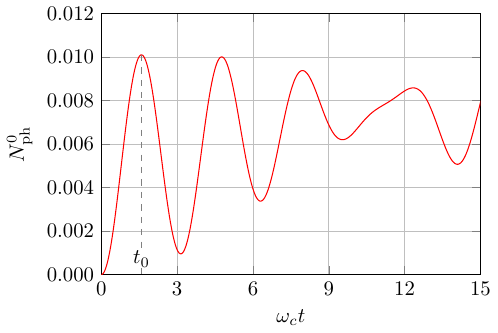}
  \caption{(Color online). The photon number $N^{0}_{\rm{ph}}$ as a
    function of $t$ in the free time evolution of the Rabi model.
    Here, we take $\omega_{c} = \omega_{a}$, $g/\omega_{c} = 0.1$ and
    $T = 15/\omega_{c}$.~\label{fig:nozero}}
\end{figure}

We consider the time evolution under the effective Hamiltonian $\hat{H}_{\rm{eff}}$ starting from the initial state $| 0 \rangle$. The average photon number $N^{0}_{\rm{ph}}$ is obtained by numerically solving the time-dependent Schr$\ddot{o}$dinger equation using exact diagonalization in a truncated subspace. The validity of this approach is confirmed by the excellent agreement with results from the nonlinear Heisenberg-picture equations derived in Appendix~\ref{app:NLDE}. Fig.~\ref{fig:nozero} shows $N^{0}_{\rm{ph}} (t)$ for $\omega_{c} = \omega_{a}$, $t \leq 15/\omega_{c}$, and ultrastrong coupling $g/\omega_{c} = 0.1$. The photon number oscillates in time and reaches its first maximum, approximately $0.01$, at $t_{0}$, which remains the maximum over the entire time interval. This demonstrates that even in the ultrastrong coupling regime, the average photon number generated under continuous evolution of $\hat{H}_{\text{eff}}$ is very small.

This raises the question of whether the average photon number can be enhanced by bang-bang control of the CRW interaction—i.e., by switching the CRW interaction on and off at appropriate times. Insights can be gained from the dynamics shown in Fig.~\ref{fig:nozero}. If the total evolution time $T \leq t_{0}$, it is reasonable to keep the CRW interaction on throughout. For $T>t_{0}$, for instance at $\omega_{c}t=3$, a higher photon number at time $3/\omega_{c}$ can be achieved by switching off the CRW interaction during the interval $0\le t\le t_{0}$ and then switching it on for the remaining time. This intuitive observation suggests that properly timed intervals without the CRW interaction can lead to a net increase in the photon number, even though no photons are generated during those periods. Our main task, therefore, is to develop numerical algorithms to identify these optimal switching schedules.

In our algorithms, we divide the total evolution time $T$ into small intervals of fixed duration $\delta t$. For each interval, we denote the choice of switching the CRW interaction on by $1$ and off by $0$. A control sequence is then represented by a binary string of length $T/\delta t$, where $n_{g}$ and $n_{0}$ denote the numbers of $1$s and $0$s, respectively, satisfying $n_{g}+n_{0}=\frac{T}{\delta t}$. Since the total number of possible sequences is $2^{T/\delta t}$, exhaustive search becomes infeasible for large $T$ and the small $\delta t$.

A potential approach is the greedy algorithm, where at each time step we choose $0$ or $1$ based on which yields a larger photon number at the end of that step. As we will show, this greedy strategy can effectively increase the average photon number.

In general, however, the greedy algorithm does not guarantee the optimal control sequence for maximizing the final photon number. To address this, we develop the pruning greedy algorithm (PGA), which constructs a significantly reduced search space through sorting and pruning. Compared with exhaustive search, the PGA allows us to identify near-optimal control sequences for larger $T$ and smaller $\delta t$. A detailed description of the PGA is provided in Appendix.~\ref{app:PGA}.

\section{Numerical results~\label{sec:bangbang}}

In this section, we present the numerical results for the average photon number obtained using the bang-bang control scheme based on the greedy algorithm and the PGA, and analyze the underlying physics.

\begin{figure}[!htbp]
  \centering \includegraphics[width=0.45\textwidth]{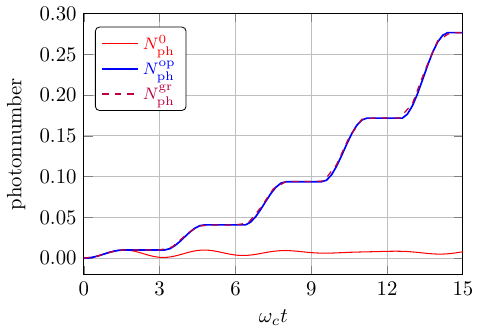}
  \caption{(Color online). The photon number $N_{\rm{ph}}$ as a
    function of $t$. The red solid line represents the photon number
    $N^{0}_{\rm{ph}}$ obtained by the free time evolution of the Rabi
    model. The blue solid line represents the photon number
    $N^{\rm{op}}_{\rm{ph}}$ obtained by the optimal quantum control of
    the CRW interaction. The purple dashed line stands for the photon number 
    $N^{\rm{gr}}_{\rm{ph}}$ obtained from the greedy algorithm. Here, we take 
    $\omega_{c} = \omega_{a}$, $g/\omega_{c} = 0.1$, $T = 15/\omega_{c}$ and 
    $\delta t = 0.2/\omega_{c}$.~\label{fig:T15}}
\end{figure}

We first demonstrate that a properly designed control sequence can significantly enhance photon generation. A typical example is shown in 
Fig.~\ref{fig:T15}, where we plot the time evolution of the average photon number,  $N_{\text{ph}}^{\text{gr}}$ and $N_{\text{ph}}^{\text{op}}$, obtained with the greedy algorithm and the PGA, respectively. The plateaus in the photon number correspond to periods during which the CRW interaction is switched off. The photon number $N_{\text{ph}}^{\text{op}}$ is slightly larger than $N_{\text{ph}}^{\text{ag}}$. For comparison, we also show the time evolution of the average photon number $N_{\text{ph}}^{0}$ under the effective Hamiltonian without control.

\begin{figure}[!htbp]
  \centering \includegraphics[width = 0.48\textwidth]{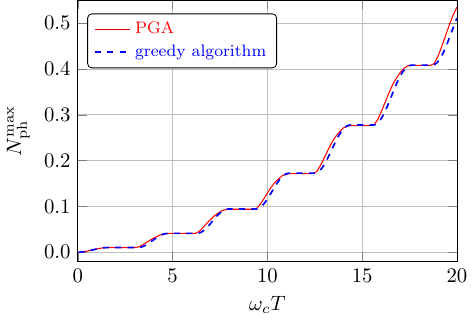}
  \caption{(Color online). The maximum photon number $N^{\rm{max}}_{\rm{ph}}$ 
  	obtained by optimal quantum control of the CRW interaction for different 
  	evolution time $T$. The red solid line represents results obtained via the 
  	pruning greedy algorithm (PGA). The blue dashed line represents results 
  	obtained by the greedy algorithm. Here, we take $\omega_{c} = \omega_{a}$, 
  	$g/\omega_{c} = 0.1$ and $\delta t = 0.2/\omega_{c}$.}
  \label{fig:g01}
\end{figure}

Figure~\ref{fig:g01} shows the average photon number as a function of the total evolution time $T$, where the results from the PGA (red solid line) and the greedy algorithm (blue dashed line) are compared. The results obtained with the PGA can be regarded as the maximum photon number $N^{\rm{max}}_{\rm{ph}}$ achievable at time $T$ under bang-bang control. In the $N^{\rm{max}}_{\rm{ph}}$ curve, a series of plateaus is observed, indicating that the CRW interaction must be switched off during these intervals to maximize the photon number. As $T$ increases, the average photon number grows, suggesting that there is no upper bound in the long-time limit, since the CRW interaction can continuously excite higher multi-photon states during the intervals between plateaus.

\begin{figure}[!htbp]
  \centering \includegraphics[width = 0.48\textwidth]{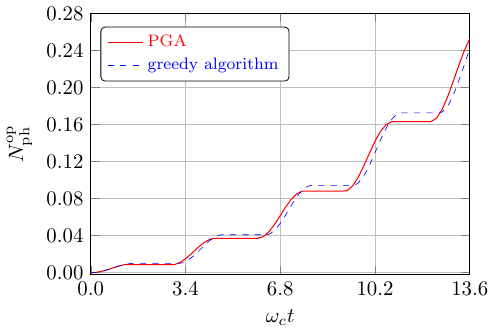}
  \caption{(Color online). The maximum photon number as a function of $t$ for 
  	$T = 13.6/\omega_{c}$. The red solid line represents the results 
  	$N^{\rm{op}}_{\rm{ph}}$ obtained by the PGA. The blue dashed line stands for 
  	the results obtained by the greedy algorithm. Here, we take $\omega_{c} = 
  	\omega_{a}$, $g/\omega_{c} = 0.1$ and $\delta t = 0.2/\omega_{c}$.}
  \label{fig:PGA}
\end{figure}

The greedy algorithm yields results only slightly lower than those of the PGA, indicating that it is effective for this problem, although it does not generally produce the optimal solution. A closer inspection of Fig.~\ref{fig:g01} reveals that the results from both algorithms nearly coincide at times corresponding to the photon number plateaus, where the control sequences are essentially the same. Discrepancies mainly occur at times between adjacent plateaus, as illustrated in Fig.~\ref{fig:PGA} for $T=13.6/\omega_{c}$.  In this case, compared with the greedy algorithm, the PGA produces plateaus that are shifted leftward in time, allowing more time in the final increasing segment, leading to a higher photon number.

\begin{figure}[!htbp]
	\centering
	\includegraphics[width = 0.475\textwidth]{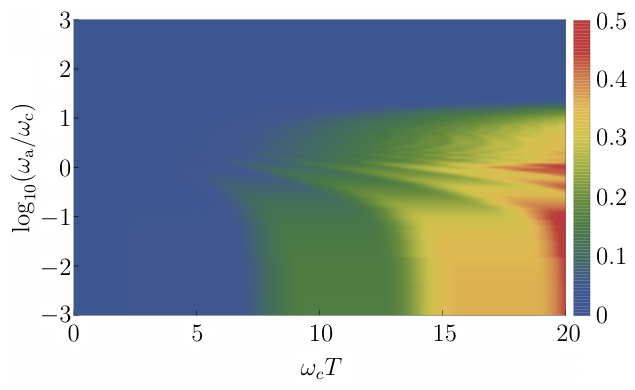}
	\caption{(Color online). The photon number $N^{\rm{gr}}_{\rm{ph}}$ obtained 
		by the greedy algorithm as a function of $T$ and $\omega_{a}$ for 
		$g/\omega_{c} = 0.1$ and $\delta t = 0.2/\omega_{c}$.}
	\label{fig:greedy2ddensity}
\end{figure}

We now extend our study to the off-resonant case where $\omega_{a}\neq\omega_{c}$. Figure~\ref{fig:greedy2ddensity} shows the photon number $N^{\rm{gr}}_{\rm{ph}}$ obtained from the greedy algorithm as a function of $T$ and $\omega_{a}$. For $\omega_{a}$ near $\omega_{c}$, different features emerge. For $\omega_{a} \gg \omega_{c}$, few photons are generated because the coupling strength $g$ is insufficient to drive transitions from the initial state $| 0 \rangle \otimes | g \rangle$ to higher energy levels. As $\omega_{a}$ decreases from $\omega_{c}$, the photon number plateaus become more pronounced and their widths increase, approaching a limiting value.

\begin{figure}[!htbp]
  \centering
  \includegraphics[width = 0.48\textwidth]{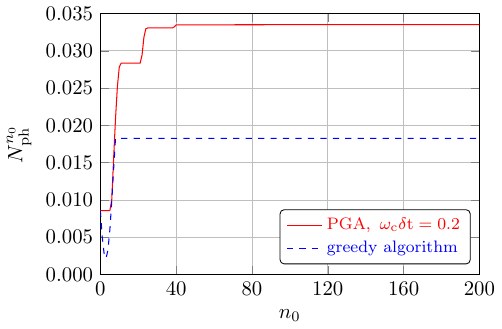}
  \caption{(Color online). The maximum photon number $N^{n_{0}}_{\rm{ph}}$ 
  	for different $n_{0}$ with the fixed $n_{g}$ and $\delta t$. The red 
  	solid line represents the results obtained by the PGA, and the blue 
  	dashed line represents the results obtained by the greedy algorithm. 
  	Here, we take $\omega_{c} = \omega_{a}$, $g/\omega_{c} = 0.1$, $n_{g} = 
  	10$ and $\delta t = 0.2/\omega_{c}$.}
  \label{fig:n0}
\end{figure}

Since photons are generated by the CRW interaction, it is natural to ask how many photons can be generated when the total time during which the CRW interaction is active is fixed. Figure~\ref{fig:n0} shows the maximal photon number $N^{n_{0}}_{\rm{ph}}$ as a function of $n_{0}$, the number of time steps with the interaction switched off, for fixed $n_{g}=10$ and $\delta t = 0.2/\omega_{c}$. The red solid line represents results from the PGA, and the blue dashed line from the greedy algorithm. According to Fig.~\ref{fig:nozero}, the total on-time of the CRW interaction is $n_{g} \delta t = 2/\omega_{c} > t_{0}$; for $n_{0}=0$, the photon number is less than $0.01$. As $n_{0}$ increases, the photon number obtained with the PGA approaches a finite limit. This limiting value characterizes the photon generation capacity of the CRW interaction when assisted by free evolution under  $\hat{H}_{0}^{\text{eff}}$ Although appropriate intervals of switching off the CRW interaction can enhance photon generation, the intrinsic properties of the CRW interaction and its total on-time ultimately determine the maximum achievable photon number.

In Fig.~\ref{fig:n0}, we also compare the greedy algorithm with the PGA. The greedy algorithm yields significantly lower photon numbers than the PGA. For small $n_{0}(<6)$, the photon number is even lower than that for $n_{0}=0$. These results indicate that the PGA is more suitable than the greedy algorithm for finding the optimal control sequence when $n_{g}$ is fixed.

\section{Conclusion and discussion~\label{sec:conclusion}}

Our numerical results demonstrate that photon generation from the vacuum can be significantly enhanced by employing a bang-bang control scheme. We emphasize that the vacuum considered here is that of the Hamiltonian $H_{0}$ rather than $H$, which draws a conceptual connection to the dynamical Casimir effect~\cite{cas48,dyncas70,Gorban2023Asymmetric}, where photons are generated by modulating the vacuum through the motion of a cavity mirror. In this sense, our bang-bang protocol for photon generation can be viewed as a generalized dynamical Casimir effect.

This perspective naturally raises the question of whether alternative bang-bang protocols exist for generating photons via the CRW interaction. We answer this in the affirmative by presenting an alternative protocol based on $\hat{\sigma}_{z}$ gates, which may be easier to implement experimentally than the switching scheme discussed above. Details of this alternative protocol are provided in Appendix~\ref{app:sigmaz}.

In the present work, we have neglected dissipation effects in our bang-bang control scheme. In general, we expect that dissipation will reduce the average photon number, an issue that is crucial for comparison with experimental realizations and warrants further investigation.

In summary, we have proposed a bang-bang control method to harness the CRW interaction for enhanced photon generation. We find that the maximum achievable photon number can be dramatically increased using the greedy algorithm and the pruning greedy algorithm. Compared with the greedy algorithm, the PGA yields better control sequences and higher photon numbers, albeit at the cost of increased computational time. We hope that our work will stimulate further studies on practical applications of the CRW interaction in diverse areas of coherent quantum manipulation.

\begin{acknowledgments}
  This work is supported by Science Challenge Project (No.TZ2025017) and the National
Key Research and Development Program of China (Grants No.2021YFA0718302 and
No.2021YFA1402104).
\end{acknowledgments}

\appendix
\begin{widetext}
\section{Dynamics of the Rabi model in the Heisenberg picture \label{app:NLDE}}

In this appendix, we develop an alternative approach to solving the dynamics of the Rabi model in the Heisenberg picture. Since the dynamics of the photon number operator $\hat{n}$ does not close within a finite set of operators, we employ a truncation method based on operator cumulants~\cite{PhysRevA.74.052110, PhysRevLett.86.568, Plankensteiner2022QuantumCumulants}.
  
It is well known that the expectation value $\langle \hat{A} \hat{B} \rangle$ can be approximated as
\begin{equation}
  \label{eq:1}
  \langle \hat{A} \hat{B} \rangle \approx \langle \hat{A} \rangle \langle \hat{B} \rangle
\end{equation}
when correlations between $\hat{A}$ and $\hat{B}$. are neglected. More generally, the expectation value of a product of $N$ operators can be expanded in terms of cumulants~\cite{PhysRevA.74.052110} as
\begin{equation}
  \left\langle \hat{O}^{1 2 \cdots N} \right\rangle \approx \sum_{\{S_{i}\}} 
   (-1)^{M} (M - 1)! \left\langle \prod_{i} \hat{O}^{(S_{i})} \right\rangle,
  \label{eq:cumulant}
\end{equation}
where $\hat{O}^{1 2 \cdots N} = \prod^{n}_{i = 1} \hat{O}^{i}$, and $M$ ($M \geq 2$) is the number of partition in the set $\{S_{i}\} = \{S_{1}, S_{2}, \cdots, S_{M}\}$ which satisfies $S_{j} \neq \varnothing$, $S_{j} \cap S_{k} = \varnothing$ and $\prod^{M}_{j = 1} \cup S_{j} = \{1, 2, \cdots, N\}$ for $\forall j, k \in \{1, 2, \cdots, M\}$.

For $N=3$ and $N=4$, Eq.~\eqref{eq:cumulant} yields
\begin{equation}
  \langle A B C \rangle \approx \langle A \rangle \langle B C \rangle + \langle B 
  \rangle \langle A C \rangle + \langle C \rangle \langle A B \rangle - 2 
  \langle A \rangle \langle B \rangle \langle C \rangle,
  \label{eq:expect_ABC}
\end{equation}
and
\begin{equation}
  \begin{split}
    \langle A B C D \rangle \approx & 6 \langle A \rangle \langle B \rangle \langle C \rangle \langle D \rangle + \langle A \rangle \langle B C D \rangle + \langle B \rangle \langle A C D \rangle + \langle C \rangle \langle A B D \rangle + \langle D \rangle \langle A B C \rangle + \langle A B \rangle
    \langle C D \rangle + \langle A C \rangle \langle B D \rangle \\
    & + \langle A D \rangle \langle B C \rangle - 2 \langle A \rangle \langle B \rangle \langle C D \rangle - 2 \langle A \rangle \langle C \rangle \langle B
    D \rangle - 2 \langle A \rangle \langle D \rangle \langle B C \rangle - 2 \langle B \rangle \langle C \rangle \langle A D \rangle - 2
    \langle B \rangle \langle D \rangle \langle A C \rangle \\
    & - 2 \langle C \rangle \langle D \rangle \langle A B \rangle.
  \label{eq:expect_ABCD}
  \end{split}
\end{equation}

Using Eq.~\eqref{eq:expect_ABC}, Vardi and Anglin derived the dynamical equations for the Bose-Einstein condensate and predicted the quantum break time~\cite{PhysRevLett.86.568}. For our problem, we find that achieving nearly accurate results requires the use of  Eq.~\eqref{eq:expect_ABCD}  to derive the nonlinear dynamical equations for the Rabi model.

We introduce the following set of dynamical operators:
\begin{subequations}
  \begin{eqnarray}
    & \hat{x} = \hat{b}^{\dagger} + \hat{b}, \hat{p} = i (\hat{b}^{\dagger} - 
      \hat{b}), \hat{n} = \hat{b}^{\dagger} \hat{b}, \hat{\gamma} = 
      (-1)^{\hat{n}}, \hat{\delta} = \dfrac{i}{2} (\hat{x} \hat{\gamma} - 
      \hat{\gamma} \hat{x}), \hat{\epsilon} = \dfrac{i}{2} (\hat{p} 
      \hat{\gamma} - \hat{\gamma} \hat{p}), \\
    & \hat{\alpha} = \hat{x}^{2}, \hat{\beta} = \hat{p}^{2}, \hat{\theta} = 
      \dfrac{1}{2} (\hat{x} \hat{p} + \hat{p} \hat{x}), \hat{\kappa} = 
      \dfrac{1}{2} (\hat{\alpha} \hat{\gamma} + \hat{\gamma} \hat{\alpha}), 
      \hat{\lambda} = \dfrac{1}{2} (\hat{\beta} \hat{\gamma} + \hat{\gamma} 
      \hat{\beta}), \hat{\mu} = \dfrac{1}{2} (\hat{\theta} \hat{\gamma} + 
      \hat{\gamma} \hat{\theta}).
  \end{eqnarray}
  \label{eq:variables}
\end{subequations}

In the Heisenberg picture, the initial conditions at $t=0$ are
\begin{equation}
    x(0) = p (0) = n (0) = \delta (0) = \epsilon (0) = \theta (0) = \lambda (0) 
    = 0, \gamma (0) = \alpha (0) = \beta (0) = \kappa (0) = \mu (0) = 1,
    \label{eq:init_condition}
\end{equation}
corresponding to the initial vacuum state.

Using Eqs.~\eqref{eq:expect_ABCD} and \eqref{eq:variables}, we obtain the following system of equations:
\begin{subequations}
  \begin{eqnarray}
      \dfrac{d x}{d t} & = & \omega_{c} p + \omega_{a} \delta, \\
      \dfrac{d p}{d t} & = & - \omega_{c} x + \omega_{a} \epsilon - 2g, \\
      \dfrac{d n}{d t} & = & - g p, \\
      \dfrac{d \gamma}{d t} & = & 2g \delta, \\
      \dfrac{d \delta}{d t} & = & \omega_{c} \epsilon - \omega_{a} x - 2g 
                             \kappa, \\
      \dfrac{d \epsilon}{d t} & = & - \omega_{c} \delta - \omega_{a} p - 2 g 
                             \lambda, \\
      \dfrac{d \alpha}{d t} & = & 2 \omega_{c} \theta, \\
      \dfrac{d \beta}{d t} & = & - 2 \omega_{c} \theta - 4 g p, \\
      \dfrac{d \theta}{d t} & = & - \omega_{c} \alpha + \omega_{c} \beta - 2 g 
                             x, \\
      \dfrac{d \kappa}{d t} & = & 2 \omega_{c} \lambda + 6 g \alpha \delta - 12 
                             g x^{2} \delta, \\
      \dfrac{d \lambda}{d t} & = & - \omega_{c} \kappa + \omega_{c} \mu + 2 g 
                             \alpha \epsilon - 4 g x^{2} \epsilon - 8 g x p \delta + 4 g \theta 
                             \delta, \\
      \dfrac{d \mu}{d t} & = & - 2 \omega_{c} \lambda + 2 g \beta \delta + 4 g 
                             \theta \epsilon - 4 g p^{2} \delta - 8 g x p \epsilon.
  \end{eqnarray}
  \label{eq:nle}
\end{subequations}

Solving these equations with the initial conditions~\eqref{eq:init_condition} for $\omega_{a} = \omega_{c}$ and $g/\omega_{c} = 0.1$ yields the photon number as a function of time, shown as the blue dashed line in Fig.~\ref{fig:nle}. The results are in excellent agreement with those obtained from exact diagonalization (red solid line), confirming the validity of our approach.
\begin{figure}[!htbp]
    \centering \subfloat[$N^{0}_{\rm{ph}} (t)$.]{
      \includegraphics[width=0.45\textwidth]{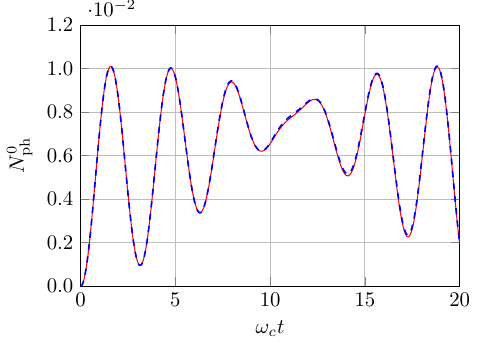}~\label{fig:nle}}
    \subfloat[$N^{\rm{op}}_{\rm{ph}} (T)$.]{
      \includegraphics[width=0.45\textwidth]{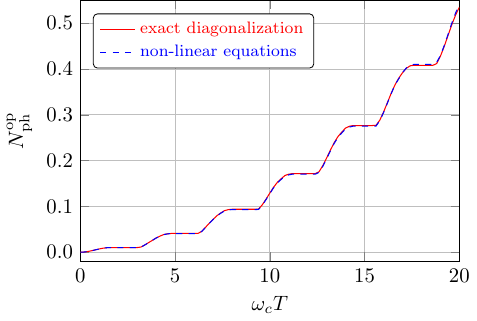}~\label{fig:nlevspga}}
    \caption{(Color online). (a) The average photon number
      $N^{0}_{\rm{ph}}$ as a function of the evolution time $t$ for
      the free time evolution of $\hat{H}$. (b) The maximum photon
      number $N^{\rm{op}}_{\rm{ph}}$ as a function of the experimental
      time $T$ for the bang-bang protocol with
      $\delta t = 0.2/\omega_{c}$. Here, we take
      $\omega_{a} = \omega_{c}$ and $g/\omega_{c} = 0.1$.}
    \label{fig:beyondMFT}
\end{figure}

We further apply this method to compute the average photon number in the bang-bang protocol, as shown in Fig.~\ref{fig:nlevspga}. The results again agree well with those from exact diagonalization, providing additional support for the numerical results presented in Sec.~\ref{sec:bangbang}.

\section{The pruning greedy algorithm \label{app:PGA}}

In this appendix, we introduce the pruning greedy algorithm (PGA), which is used to search for the optimal control sequence in our problem.

  For a fixed evolution time $T$ and a small time step $\delta t$, each control sequence is represented by a binary string of length $T/\delta t$. The total number of possible sequences is $2^{T/\delta t}$, making exhaustive search impractical for obtaining the optimal sequence.

  The PGA combines elements of exhaustive search and the greedy algorithm. We introduce an integer 
 $N$ ($1\le N \leq T/\delta t$) that defines the size of the truncated search subspace, which contains at most $2^{N}$ sequences.

  The procedure of the PGA is as follows. If $T/\delta t\le N$, we perform an exhaustive search to obtain the optimal sequence. If $T/\delta t> N$, we reduce the search space to a subspace of size $2^{N}$ and then search exhaustively within it. To construct this subspace, we begin at time $t=N \delta t$, where the search space contains $2^{N}$ sequences. When advancing to time $t=(N+1) \delta t$, the number of candidate sequences grows to $2^{N+1}$. We then truncate it back to $2^{N}$ by retaining only the $2^{N}$ sequences with the largest average photon numbers at that time. This truncation process is repeated until $t=T$, yielding a search subspace of size $2^{N}$ at the final time.

  Note that the PGA reduces to the greedy algorithm when $N = 1$ and to exhaustive search when
  $N = T/\delta t$. The convergence of the PGA can be ensured by choosing a sufficiently large $N$.

  In essence, our problem amounts to finding the optimal path in a binary tree, where we iteratively prune the tree to keep the subspace manageable. Hence, we term this method the “pruning greedy algorithm.” The pseudo-code is provided in Algorithm.~\ref{algPGA}.
  
  \begin{algorithm}[!htbp]
    \SetNlSkip{0.25em} \SetInd{0.5em}{1em} \KwIn{the evolution time
      $T$, the pulse duration $\delta t$, the truncated number $N$}
    \KwOut{the maximum photon number $N^{\rm{op}}_{\rm{ph}}$ with its
      corresponding control sequence} \eIf{$T/\delta t \le N$} {output the
      maximal photon number at time $T$ and its control sequence by
      exhaustive searching\;} {constructing the searching subspace
      with $2^{N}$ possible sequences at time $t_{L} = L \delta t$ with
      $L=N$ \; \While{$L < T/\delta t$} {add $0$ or $1$ as the next number
        for every binary sequence in the previous searching subspace
        \; sorting these $2^{N + 1}$ sequences in the decreasing order
        by the corresponding photon numbers \; keep the first $2^{N}$
        sequences as the new searching subspace \; $L = L + 1$ \;}
      output the maximal photon number at time $T$ and its control
      sequence}
    \caption{The pseudo-code of the `pruning greedy algorithm'}
    \label{algPGA}
  \end{algorithm}

  Our algorithm discretizes the optimal quantum control problem. To ensure convergence to the continuous-time limit, the pulse duration $\delta t$ must be sufficiently small. Fig.~\ref{fig:PGA} shows the maximum photon number $N^{\rm{max}}_{\rm{ph}}$ obtained
  by PGA with $\delta t = 0.1/\omega_{c}$ and $\delta t = 0.2/\omega_{c}$ as a function of the evolution time $T$. The two curves are in good agreement, confirming the convergence of our results.
  
  \begin{figure}[!htbp]
    \centering \includegraphics[width=0.6\textwidth]{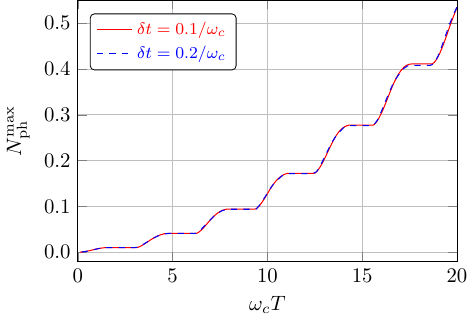}
    \caption{(Color online). The maximum photon number
      $N^{\rm{max}}_{\rm{ph}}$ as a function of the experimental time
      $T$ for the bang-bang protocol. The red solid line and blue
      dashed line represent the results obtained by the PGA with
      $\delta t = 0.1/\omega_{c}$ and $\delta t = 0.2/\omega_{c}$, respectively. Here, we
      take $\omega_{a} = \omega_{c}$ and $g/\omega_{c} = 0.1$.}
    \label{fig:pgadt}
  \end{figure}

\section{Bang-bang control via $\sigma_z$ gates \label{app:sigmaz}}

  In this appendix, we present an alternative bang-bang control scheme for generating photons via the CRW interaction using $\hat{\sigma}_{z}$ gates, which may be more amenable to experimental implementation than the scheme based on directly switching the coupling on and off.

  The protocol based on $\hat{\sigma}_{z}$ gates is conceptually similar to the on–off switching scheme. Instead of switching off the CRW interaction, we apply two $\hat{\sigma}_{z}$ gates at the beginning and the end of a given time interval. This effectively replaces the free Hamiltonian $\hat{H}_{0}$ with
  \begin{equation}
    \hat{H}' = \hat{\sigma}_{z} \hat{H} \hat{\sigma}_{z} = \omega_{c} 
    \hat{a}^{\dagger} \hat{a} + \dfrac{\omega_{a}}{2} \hat{\sigma}_{z} - g 
    \hat{\sigma}_{x} (\hat{a}^{\dagger} + \hat{a}).
  \end{equation}

  \begin{figure}[!htbp]
    \centering
    \subfloat[$N^{\rm{max}}_{\rm{ph}} (T)$.]{\includegraphics[width =
      0.425\textwidth]{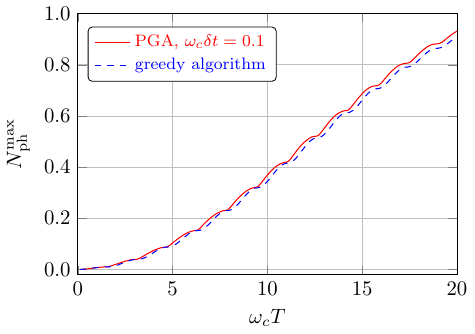}~\label{fig:gmgop}}
    \subfloat[$N^{\rm{gr}}_{\rm{ph}} (T,
    \omega_{a})$]{\includegraphics[width =
      0.5\textwidth]{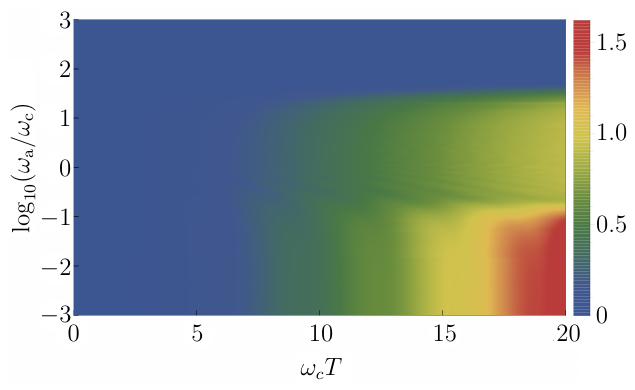}~\label{fig:gmgdensity}}
    \caption{(Color online). (a) The maximum photon number
      $N^{\rm{max}}_{\rm{ph}}$ as a function of the experimental time
      $T$ for $\omega_{a} = \omega_{c}$. (b) The photon number
      $N^{\rm{gr}}_{\rm{ph}}$ obtained by the greedy algorithm as a
      function of $T$ and $\omega_{a}$. Here, we take
      $g/\omega_{c} = 0.1$ and $\delta t = 0.1/\omega_{c}$.}
    \label{fig:gmg}
  \end{figure}

  Using the PGA described in Appendix~\ref{app:PGA} and the greedy algorithm, we compute the maximum photon number $N^{\rm{max}}_{\rm{ph}}$ as a function of the evolution time $T$, with the results shown in Fig.~\ref{fig:gmgop}. The achieved photon numbers are significantly larger than those obtained from free evolution under the Rabi model. Moreover, the results from the greedy algorithm are close to those from the PGA, suggesting that the greedy approach is also experimentally useful. 
  
  We further evaluate the maximum photon number $N^{\rm{gr}}_{\rm{ph}}$ for various  $\omega_{a}$ using the greedy algorithm. As shown in Fig.~\ref{fig:gmgdensity}, for a fixed $T$, $N^{\rm{gr}}_{\rm{ph}}$ increases as $\omega_{a}$ decreases, while it approaches zero in the regime $\omega_{a} \gg \omega_{c}$.

\end{widetext}

\bibliography{CRWref}

\end{document}